\documentclass[aps,pra,reprint,groupedaddress,showpacs,nofootinbib]{revtex4-1}
\usepackage{graphicx,amsmath,amssymb,amsfonts,epstopdf,booktabs,dcolumn}
\usepackage{makecell,tikz}
\newcommand{\ket}[1]{|#1\rangle}
\newcommand{\bra}[1]{\langle{#1}|}
\newcommand{\braket}[1]{\langle{#1}\rangle}

\newcolumntype{x}[1]{>{\centering\arraybackslash}p{#1}}
\newcommand\diag[4]{%
  \multicolumn{1}{p{#2}|}{\hskip-\tabcolsep
  $\vcenter{\begin{tikzpicture}[baseline=0,anchor=south west,inner sep=#1]
  \path[use as bounding box] (0,0) rectangle (#2+2\tabcolsep,\baselineskip);
  \node[minimum width={#2+2\tabcolsep},minimum height=\baselineskip+\extrarowheight] (box) {};
  \draw (box.north west) -- (box.south east);
  \node[anchor=south west] at (box.south west) {#3};
  \node[anchor=north east] at (box.north east) {#4};
 \end{tikzpicture}}$\hskip-\tabcolsep}}

\begin{document}

\title{Experimental determination of the nuclear magnetic octupole moment of $^{137}$Ba$^+$ ion}

\author{Nicholas C. Lewty}
\email{nicholas.lewty@nus.edu.sg}
\affiliation{Centre for Quantum Technologies and Department of Physics, National University of Singapore, 3 Science Drive 2, 117543 Singapore}
\author{Boon Leng Chuah}
\affiliation{Centre for Quantum Technologies and Department of Physics, National University of Singapore, 3 Science Drive 2, 117543 Singapore}
\author{Radu Cazan}
\affiliation{Centre for Quantum Technologies and Department of Physics, National University of Singapore, 3 Science Drive 2, 117543 Singapore}
\author{B. K. Sahoo}
\affiliation{Theoretical Physics Division, Physical Research Laboratory, Ahmedabad-380009, India}
\author{Murray D. Barrett}
\affiliation{Centre for Quantum Technologies and Department of Physics, National University of Singapore, 3 Science Drive 2, 117543 Singapore}

\date{\today}

\begin{abstract}We perform precision measurements on the 5D$_{5/2}$ manifold hyperfine intervals of a single trapped ion, $^{137}$Ba$^+$. RF spectroscopy is used to measure the hyperfine intervals to an accuracy of a few Hz.  Our results provide a three orders of magnitude improvement in accuracy over previous work and also provide a 10-fold improvement in the value of $g_J$ for this level. These results complement our previous work on the 5D$_{3/2}$ manifold of $^{137}$Ba$^+$, providing an independent measurement of the nuclear octupole, and a consistency check on atomic structure calculations.
\end{abstract}
\maketitle


High precision measurements of the hyperfine structure provide stringent tests for state-of-the-art atomic structure calculations. Comparing measured hyperfine structure constants with calculated values allows one to experimentally assess the accuracy of the structure calculations \cite{Sahoo2006}. These calculations play a crucial role in the interpretation of parity nonconservation (PNC) experiments which provide important tests of the standard model at low energy \cite{Langacker1992}. In addition, the hyperfine structure provides insight into the nuclear structure of atoms \cite{Arimondo}.  By using standard ion manipulation techniques \cite{Dietrich2010,Chuah2011} and high precision rf spectroscopy methods \cite{Koerber2002} on a single trapped ion, the hyperfine intervals of the 5D$_{5/2}$ ($\tau \approx 32$s \cite{Gurell2007}) manifold are measured to an accuracy of a few Hz. These results complement our previous work on the 5D$_{3/2}$ manifold of $^{137}$Ba$^+$ (I = 3/2), providing an independent measurement of the nuclear octupole, and a consistency check on the associated atomic structure calculations.

Using the notation from \cite{Beloy2008b}, the zero field hyperfine intervals $\delta W_F = W_F-W_{F+1}$ of the 5D$_{5/2}$ manifold are
\begin{align}
\delta W^{(5/2)}_1 &=-2\textrm{A}+\frac{4}{5}\textrm{B}-\frac{96}{5}\textrm{C}-\frac{1}{75}\eta, \label{split1}\\
\label{split2}
\delta W^{(5/2)}_2 &= -3\textrm{A}+\frac{9}{20}\textrm{B}+\frac{81}{5}\textrm{C}-\frac{1}{300}\eta -\frac{1}{20}\sqrt{\frac{3}{7}}\zeta,\\
\label{split3}
\delta W^{(5/2)}_3 &=-4\textrm{A}-\frac{4}{5}\textrm{B}-\frac{32}{5}\textrm{C}+\frac{2}{75}\eta+\frac{2}{25\sqrt{21}}\zeta,
\end{align} where $\eta$ and $\zeta$ are the second-order correction terms characterizing the hyperfine mixing with the lower 5D$_{3/2}$ manifold. Detailed expressions for these terms are given in Appendix A. By solving the above equations one can get the hyperfine constant C from measurements of the hyperfine intervals, $\delta{W}^{(5/2)}_F$, and the correction factor, $\zeta$, via the equation
\begin{align}
\label{C52}
\textrm{C}(5D_{5/2}) =& - \frac{1}{40}\delta W^{(5/2)}_{1} + \frac{1}{35}\delta W^{(5/2)}_{2} \nonumber \\
&- \frac{1}{112}\delta W^{(5/2)}_{3} + \frac{1}{200 \sqrt{21}} \zeta.
\end{align} From atomic structure calculations \cite{Sahoo2006,Sahoo2006a}, the nuclear magnetic octupole moment $\Omega(^{137}\mathrm{Ba}^{+}_{\,\mathrm{D_{5/2}}})$ can be related to the hyperfine constant $\textrm{C}(5D_{5/2})$ via

\begin{align}
\label{oct52}
\textrm{C}(5D_{5/2}) = - 0.25(1)\left( \frac{\Omega}{\mu_N \times \rm{b} }\right)\,\rm{kHz},
\end{align} where $\mu_N$ is the Bohr magneton and b is the barn unit of area.

In the 5D$_{3/2}$ manifold the equivalent equations are
\begin{align}
\label{C32}
\textrm{C}(5D_{3/2}) =& - \frac{1}{80}\delta W^{(3/2)}_{0} + \frac{1}{100}\delta W^{(3/2)}_{1} \nonumber \\
&- \frac{1}{400}\delta W^{(3/2)}_{2} - \frac{1}{2000 \sqrt{21}} \zeta
\end{align}
and
\begin{equation}
\label{oct32}
\textrm{C}(5D_{3/2}) = 0.584(6)\left( \frac{\Omega}{\mu_N \times \rm{b} }\right)\,\rm{kHz},
\end{equation} which provides an independent determination of the octupole moment and a consistency check in the associated atomic structure calculations. Combining the measurements of the hyperfine C constant in the 5D$_{5/2}$ manifold with the previous measurements in the 5D$_{3/2}$ manifold \cite{Lewty2012}, the correction factor, $\zeta$, can be eliminated resulting in equations
\begin{align}
&\textrm{C}(5D_{3/2}) + \frac{1}{10}\textrm{C}(5D_{5/2}) = \nonumber \\
& - \frac{1}{80}\delta W^{(3/2)}_{0} + \frac{1}{100}\delta W^{(3/2)}_{1} - \frac{1}{400}\delta W^{(3/2)}_{2} \nonumber \\
&- \frac{1}{400}\delta W^{(5/2)}_{1} + \frac{1}{350}\delta W^{(5/2)}_{2} - \frac{1}{1120}\delta W^{(5/2)}_{3},
\end{align}
and
\begin{align}
\label{omega}
\textrm{C}(5D_{3/2}) + \frac{1}{10}\textrm{C}(5D_{5/2}) = \nonumber \\ 
0.559(6)\left( \frac{\Omega}{\mu_N \times \rm{b} }\right)\,\rm{kHz}.
\end{align}

\begin{figure}
\centering
\includegraphics[width=0.5\textwidth]{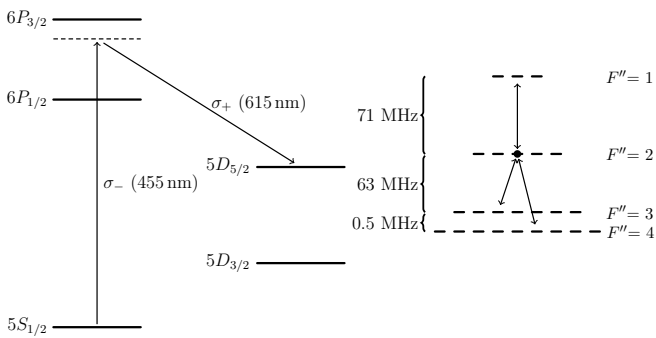}
\caption{Level structure for rf spectroscopy of 5D$_{5/2}$ manifold: The ion is prepared in the $\left|F=2, m_F=2\right\rangle$ state of the 6S$_{1/2}$ level from where it is shelved to $\left|F''= 2, m_{F''}=0\right\rangle$ of the 5D$_{5/2}$ level with a pair of Raman beams which are red-detuned by $\approx \Delta=2\pi\times 20\,\mathrm{GHz}$ from the 6P$_{3/2}$ level. The rf transition is detected by shelving back to the 5S$_{1/2}$ level and fluorescing (see text). The three rf transitions of interest are indicated by the arrows.}
\label{levelstructure}
\end{figure}


To determine the hyperfine constant C, we infer $\delta W^{(5/2)}_k$ from measurements of the splittings over a range of finite magnetic fields and extrapolating the results to zero field. For such measurements, states least sensitive to magnetic fields are preferred as they allow for longer integration times and hence higher precision. Due to the small energy difference  of $\approx$ $500\, \mathrm{kHz}$ between $F''=3$ and $F''=4$ levels of the 5D$_{5/2}$ manifold there is a significant amount of Zeeman mixing between these states even for low magnetic fields.  As a consequence, the second order zeeman shifts of the $\ket{F''=3,m_{F''}=0}$ and $\ket{F''=4,m_{F''}=0}$ energies are large and therefore the accuracy in determining the intervals is greatly limited by magnetic field fluctuations. Instead it is much more favorable to use states $\ket{\pm}$, which, neglecting mixing with the $F''=1,\mbox{ and }F''=2$ states, have the approximate form
\begin{align}
 \ket{+} =& \sin{\theta_+}\ket{F''=3,m_{F''}=+1} \nonumber \\
 &+ \cos{\theta_+}\ket{F''=4,m_{F''}=+1},
\end{align}
\begin{align}
 \ket{-} =& \cos{\theta_-}\ket{F''=3,m_{F''}=-1} \nonumber \\
 & - \sin{\theta_-}\ket{F''=4,m_{F''}=-1}
\end{align} for positive magnetic fields, as proposed in \cite{Beloy2008a}.  The mixing angles $\theta_\pm$ are functions of the applied magnetic field with $\theta_\pm=0$ at zero field and $\theta \approx \frac{\pi}{4}$ over the magnetic field range of $0.4-2\,\mathrm{G}$ explored in this work. Over the same magnetic field range, $\ket{1}\equiv\ \ket{F''=1,m_{F''}=0}$, and $\ket{2}\equiv\ket{F''=2,m_{F''}=0}$ are also only weakly dependent on the magnetic field.  Therefore measurements are carried out on the transitions $\ket{2}\leftrightarrow\ket{1}$ and $\ket{2}\leftrightarrow\ket{\pm}$ as depicted in Fig. \ref{levelstructure}. It is worth noting that the $\ket{2} \leftrightarrow \ket{-}$ transition contains two turning points in its dependence on the magnetic field and the  $\ket{2} \leftrightarrow \ket{+}$ transition contains one. As with our previous measurement of the 5D$_{3/2}$ hyperfine intervals \cite{Lewty2012}, the full magnetic field dependence of the transitions is mapped out.

\begin{figure*}
\centering
\includegraphics[width=1\textwidth]{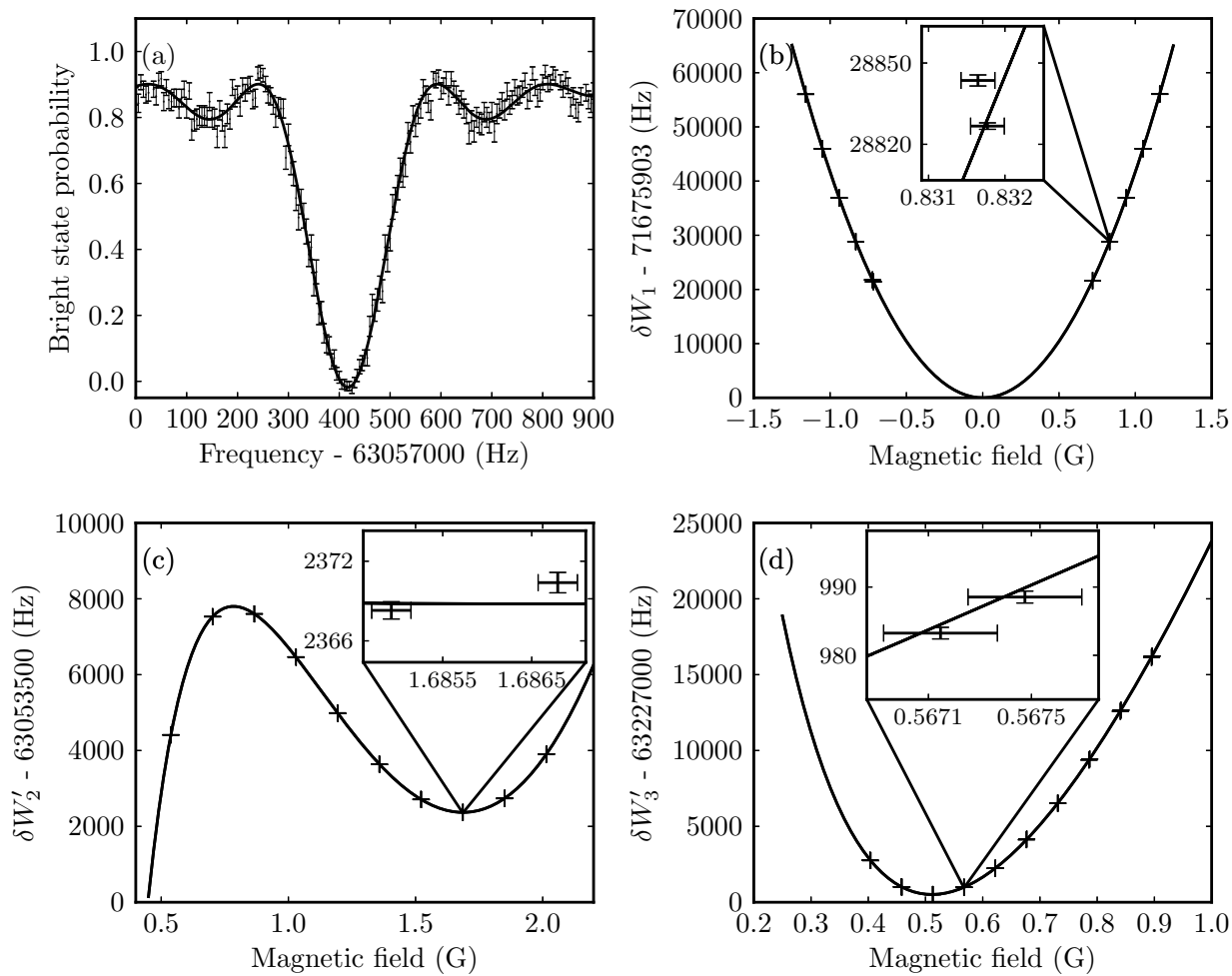}
\caption{(a) Transition probability for ion being in bright state at a magnetic field of $B\approx 1\,\mathrm{G}$ as a function of the external rf frequency for the $\ket{F''=2, m_{F''}=0}$ $\leftrightarrow$ $\ket{F''=3, m_{F''}=-1}$ transition. Plots of measured hyperfine intervals as a function of magnetic field for (b) $\delta{W}_{1} = $ $\ket{2} \leftrightarrow \ket{1}$, (c) $\delta{W}^\prime_{2} = $ $\ket{2} \leftrightarrow \ket{-}$ and (d) $\delta{W}^\prime_{3} = $ $\ket{2} \leftrightarrow \ket{+}$. The crosses represent the measured data and the solid black lines are fits to the data (See text for further details).}
\label{plots}
\end{figure*}

The method for measuring the hyperfine intervals of the 5D$_{5/2}$ manifold is similar to one detailed in \cite{Lewty2012}. Briefly, the ion is first Doppler cooled and then optically pumped to the $\ket{F = 2,m_F = 2}$ ground state. The ion is then shelved to $\ket{2}$ state with $\approx$ 95\% efficiency using a two color Raman transition with $455\, \mathrm{nm}$ and $615\, \mathrm{nm}$ lasers. An rf antenna, driven by a signal generator synchronized to a GPS-disciplined Rb clock, is then used to drive the transition of interest. To determine if the hyperfine transition occurred we use a second Raman pulse to transfer the ion from the $\ket{2}$ state back to the $\ket{F=2,m_F=2}$ ground state. The ion fluorescence from the Doppler cooling lasers is then counted using a single photon counting module. Thus, the fluorescence counts provide a probability measurement of the rf transition taking place: the ion being dark if the rf transition took place and bright otherwise. This process is repeated 200 times to obtain enough statistical data for one frequency point of an rf resonance scan.

The rf resonance scans are taken by stepping the rf signal generator in $5\, \mathrm{Hz}$ steps. The rf signal power is adjusted until the $\pi$-pulse time for the resonant transition is approximately $5\,\mathrm{ms}$.  The scans are fitted via a $\chi^2$ minimization \cite{Marquardt1963} to the usual Rabi flopping function \cite{Metcalf1999} with additional offset and amplitude parameters to account for imperfect shelving. An example of a fitted scan is shown in Fig. \ref{plots}(a). Scans are repeated over a range of magnetic fields to map out the field dependence of the measured hyperfine interval.

The magnetic field for each rf resonance scan is measured in the 5D$_{3/2}$ manifold using the $\ket{F''= 2, m_{F''} = 0} \leftrightarrow \ket{F''=3,m_{F''}=\pm1}$ transitions as in our previous work \cite{Lewty2012}. The time lapsed between taking the first magnetic field calibration measurement and measuring the hyperfine interval of interest is about 10 mins. It is therefore necessary to consider the effect of slow magnetic field drifts in this time period as these give rise to additional errors in the magnetic field calibration. The magnetic field drift was measured over a period of two hours to generate sufficient statistical data on the RMS of this slow drift. We determine the magnetic field to have 220$\,\mathrm{\mu{G}}$ RMS drift over a 10 minute interval. This is a factor of four larger than the error in the fit of the magnetic field resonance scan.

The systematic influence of micromotion, stray repump light, stray Raman light and off resonant rf coupling encountered in this measurement are detailed in \cite{Lewty2012}. These errors do not have a large impact here as they are less than $0.5\,\mathrm{Hz}$, which is smaller than the statistical error on the measured hyperfine intervals. However, there is an additional systematic error in the 5D$_{5/2}$ hyperfine interval measurements which is not present in the  5D$_{3/2}$ hyperfine interval measurements. The cause of this additional error originates from the proximity of the $F''= 3$ and $F''=4$ levels. The rf field driving the trap electrodes induce currents that generate an rf magnetic field in the trap. The rf drive frequency is sufficiently close to transition frequencies between the $m_{F''}$ states of the $F''= 3$ and $F''=4$ levels that a significant ac-Zeeman shift results. This effect has previously been observed with trapped ions in \cite{Sherman2008}.

In the ideal geometry the null in the electric field would occur at the center of the trap and magnetic fields from the induced currents would cancel. However, fabrication imperfections or design asymmetries result in a non-zero magnetic field at the ion's position. To estimate the size of this field, we assume the rf electrodes carry equal currents.  This gives rise to a zero magnetic field at the midpoint between the two electrodes. For a small displacement, $\delta$, from the midpoint position, the magnitude of the $B$-field is then given by
\begin{equation}
B=\frac{8 \mu_0 I}{2\pi a^2}\delta = \frac{8 \mu_0 V_0 \Omega C}{2\pi a^2}\delta
\end{equation}
where $a$ is the separation of the electrodes, $V_0$ and $\Omega$ are the amplitude and frequency of the rf trapping potential, respectively, and $C$ is the electrode capacitance.  In our system we have $a=1.77\,\mathrm{mm}$, $V_0 = 120\,\mathrm{V}$, and $\Omega=2\pi\times10.6\,\mathrm{MHz}$.  For our trap dimensions, the electrode capacitance is estimated to be $C\approx 10\,\mathrm{pF}$ and, using $\delta=25\,\mathrm{\mu m}$ as a typical dimensional tolerance, we estimate a $B$-field amplitude of approximately $10\,\mathrm{mG}$.

For a given polarization and magnitude of the rf magnetic field, it is straightforward to calculate the ac-Zeeman shifts of each level as a function of the applied static magnetic field. To a good approximation this can be done by neglecting mixing with $F''=1$ and $F''=2$ levels as done in \cite{Beloy2008a}. For the estimated rf field amplitude of $10\,\mathrm{mG}$, the resulting ac-Zeeman shift can be on the order of $10\,\mathrm{Hz}$ and thus must be accounted for. 

\begin{figure}
\centering
\includegraphics[width=0.5\textwidth]{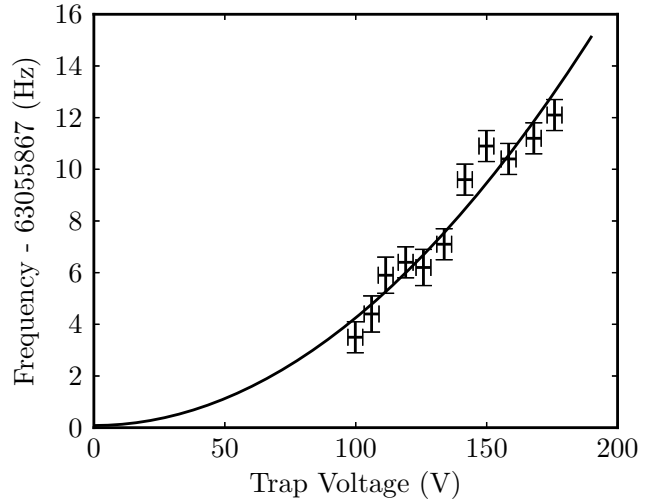}
\caption{ac-Zeeman shift measured against applied trap potential at an rf drive frequency of $10.6\,\mathrm{MHz}$ and a magnetic field of $B= 1.685\,\mathrm{G}$}
\label{stark}
\end{figure}

In order to precisely measure the ac-Zeeman shift at a particular static field it is necessary to measure the hyperfine transitions for a range of rf drive strengths and extrapolate the results to $V_0=0$. However, this is only practical at the field independent points where magnetic field drifts do not shift the resonance frequency.  Measurements for the field independent point in the $\ket{2} \leftrightarrow \ket{-}$ transition at $1.68\,\mathrm{G}$ are illustrated in Fig. \ref{stark}. Due to the fact that the ac-Zeeman shift scales with the square of the $B$-field and thus with the square of the drive voltage $V_0$, a quadratic fit centered at a rf drive voltage $V_0 = 0$ is used. Similar measurements at the other field independent points were inconclusive as the transition frequency did not significantly shift over the range of possible rf drive voltages.

Since the ac-Zeeman shifts can only be reliably measured at one magnetic field independent point, we scale the calculated values to coincide with the measured value.  In principle this requires an estimation of the polarization components of the rf magnetic field. Although the components cannot be readily determined, for a given field strength the ac-Zeeman shifts due to a $\pi$ polarized field are about two orders of magnitude smaller than for the $\sigma^\pm$ components.  For the expected field amplitude of $10\,\mathrm{mG}$, the shifts are less than $0.2\,\mathrm{Hz}$ and thus we can safely neglect the $\pi$ component.  Furthermore, a significant difference in the amplitude of the $\sigma^+$ and $\sigma^-$ components can only be present when there is a phase shift in the currents and thus voltages between the rf rods. This phase shift would result in micromotion that cannot be compensated. However, from the micromotion compensation level achieved in this trap \cite{Chuah2013}, we can safely neglect any phase shift and assume equal contributions from the $\sigma^+$ and $\sigma^-$ components.  We can thus scale the calculated ac-Zeeman shifts to the value obtained for the field independent point in the $\ket{2} \leftrightarrow \ket{-}$ transition at $1.68\,\mathrm{G}$.  We note that the estimated rf $B$-field amplitude based on this approach is $10.8(6)\,\mathrm{mG}$, which is consistent with the crude estimate above.  We also note that the inferred ac-Zeeman shifts for the other field independent points are approximately $2\,\mathrm{Hz}$ or less. This is on the order of the error of the rf resonance scan and is thus consistent with the inconclusive measurements at these points. From the ac-Zeeman shift calculations and the inferred rf $B$-field amplitude, the data points are corrected to remove the ac-Zeeman shift.

The full magnetic field dependence of the three hyperfine intervals is plotted in Figs. \ref{plots} (b), (c) and (d). For each hyperfine interval the rf transition frequency is measured at ten magnetic field values. The magnetic field points have roughly equal spacing with the range of values depending on which hyperfine interval is measured.  Two data sets for each hyperfine interval were taken on separate days and the results are given in Figs. \ref{plots} (b), (c) and (d). The insets show two data points taken on separate days to highlight the error bars. The vertical error bars represents 68\% confidence intervals from the resonance scan fits. The horizontal error bars include the magnetic field drift plus the 68\% confidence intervals from the magnetic field measurement scans.

Using a $\chi^2$ minimization, we perform a global fit of all measured hyperfine intervals and magnetic fields to the appropriate eigenvalues of the Zeeman Hamiltonian given in Appendix B. The model has a total of 5 parameters: the three zero field hyperfine splittings, $\delta W_F$, the Land\'{e} g-factor, $g_J$, and the mixing coefficient, $\beta_3$, that takes into account the effect of hyperfine mixing between the $F=3$ levels of the 5D$_{5/2}$ and 5D$_{3/2}$ manifolds. To a good approximation this mixing only effects the energies of the $\ket{\pm}$ states as is discussed in Appendix B. Due to the sensitivity of the fit to $g_J$, we included it as a fitting parameter. For the mixing coefficient, we note that since $\beta_3$ depends on exactly the same matrix elements as the correction factors $\eta$ and $\zeta$, the coefficient $\beta_3$ is not independent of $\eta$ and $\zeta$.  Moreover, as a function of the static $B$-field, both the ac-Zeeman shift and the hyperfine mixing give approximately linear shifts in the energies of the $\ket{\pm}$ states. Thus an error in one can be compensated to a degree by a change in the other.  For these reasons we leave $\beta_3$ fixed to the theoretical value of $\beta_3=1.698(17)\times 10^{-5}$.

Fitting the experimental data gives a reduced $\chi^2=1.10$ and the resulting zero field values are given in Table~\ref{fval}. The statistical errors given are the 68\% confidence intervals extracted from the fits using standard statistical methods. The systematic error accounts for the uncertainty in the theoretical value for the $\beta_3$ parameter and the uncertainty in the measured ac-Zeeman shift. The derived systematic errors in the zero field splittings and $g_J$ correspond to the largest change in the results of the fit when changing the input parameters by their estimated one standard deviation error. For $g_J$ we obtain a value of 
\begin{equation}
g_J = 1.20057(5_{\,\mathrm{Stat}})(2_{\,\mathrm{Syst}}),
\end{equation} which is within 3$\sigma$ of the value reported in \cite{Kurz2010}, but is one order of magnitude more accurate. During the preparation of this manuscript we became aware that the authors of \cite{Kurz2010} made an improved measurement of $g_J$ \cite{Hoffman2013}, which provides a 50-fold improvement in accuracy over their previous work. We note that our value is also within 3$\sigma$ of their new value.  We also note that, if $\beta_3$ is included as a fitting parameter, the fitted value is 13\% smaller than the theoretical one and the estimated zero field splittings do not change by more than twice the total error (statistical+systematic) with the most significant change occurring for $\delta W_3$.

\begin{table}[ht]
\centering\caption{\label{fval} Measured hyperfine intervals, $\delta W_k$, for the 5D$_{5/2}$ manifold of $^{137}$Ba$^{+}$.}
\begin{ruledtabular}
\begin{tabular}{lccc}
Transition & $\delta W$ (Hz) & Stat err. & Syst err. \\
\hline F=1 $\rightarrow$ F=2 & 71675902.4 & $\pm$4.6 & $\pm$0.9  \\
F=2 $\rightarrow$ F=3 & 62872301.0 & $\pm$1.4 & $\mp$2.2 \\
F=3 $\rightarrow$ F=4 &  503510.5 & $\pm$2.6 & $\pm$3.2 \\
\end{tabular}
\end{ruledtabular}
\end{table}

\begin{table}[ht]
\centering\caption{\label{hfc}D$_{5/2}$ Hyperfine coupling constants.}
\begin{ruledtabular}
\begin{tabular}{lrrr}
          & A\,(Hz)           & B\,(Hz)          & C\,(Hz) \\
\hline Uncorr.       & $-$12029724.1(9) & 59519566.2(43) & $-$41.73(18) \\
$\eta$ corr.  &          537(11) &       5367(110) & $-$ \\
$\zeta$ corr. &      $-$46.9(12) &        587(15) & 29.33(75) \\
Corr.         &    $-$12029234(11) &    59525520(110) & $-$12.41(77)\\
\end{tabular}
\end{ruledtabular}
\end{table}

The hyperfine coupling constants are determined from the obtained hyperfine intervals and are presented in Table \ref{hfc}. The statistical and systematic errors on the hyperfine intervals are added to give the total error on the hyperfine constants. Each constant falls within one standard deviation of previous work \cite{Silverans1986} but all are three orders of magnitude more accurate. The two measurements of the hyperfine constants C performed in the 5D$_{3/2}$ and 5D$_{5/2}$ manifolds allow the nuclear octupole moment of $^{137}\mathrm{Ba}^{+}$ to be calculated three different ways. From Eqs.~(\ref{oct52}), (\ref{oct32}) and (\ref{omega}) we have
\begin{equation}
\label{32_octupole}
\Omega(^{137}\mathrm{Ba}^{+}_{\,\mathrm{D_{3/2}}})=0.05057(54)~(\mu_\mathrm{N} \times \rm{b}),
\end{equation}

\begin{equation}
\label{52_octupole}
\Omega(^{137}\mathrm{Ba}^{+}_{\,\mathrm{D_{5/2}}})=0.0496(37)~(\mu_\mathrm{N} \times \rm{b}),
\end{equation}

\begin{equation}
\label{combined_octupole}
\Omega(^{137}\mathrm{Ba^{+}})=0.05061(56)~(\mu_\mathrm{N} \times \rm{b}),
\end{equation} which are all within one standard deviation of each other. They also act as a self consistency test between the two separate measurements. As all three values lie within one confidence interval, the perturbation theory used to obtain $\zeta$ in \cite{Beloy2008} and the off diagonal matrix elements given in Appendix A proves to be accurate. Do note that the sign of the octupole moment has changed with respect to our previously published value \cite{Lewty2012}, which is due to a change in convention of how we relate the octupole moment to the hyperfine C constant. This is now consistent with the description given in \cite{Schwartz1955}. Our calculations for the off diagonal matrix elements have improved with respect to \cite{Lewty2013} and for this reason the D$_{3/2}$ hyperfine constants are reviewed in Appendix C.

In summary, we have performed simultaneous high precision measurements of the hyperfine splittings of the 5D$_{3/2}$ and 5D$_{5/2}$ manifolds of $^{137}\mathrm{Ba}^{+}$, that provide an independent measurement of the nuclear octupole moment and a self consistency check of the associated structure calculation. We have also provided an improved value of g$_J$ for the 5D$_{5/2}$ manifold, which has a 10 fold improvement in accuracy.  Our measurements have sufficient precision that hyperfine mixing between the two fine structure levels must be taken into account.  Although we are only sensitive to mixing of the $F=3$ levels, we note that measurements of the splittings in the 5D$_{3/2}$ manifold for transitions between $m_F=\pm 1$ states would also be dependent on the mixing of the $F=1 \mbox{ and }2$ levels.  Thus, in principle, our measurements could be improved such that all three mixing coefficients, $\beta_k$, become measurable quantities.  This would provide a direct measurement of the reduced matrix elements $\bra{D_{3/2}}|T^e_k|\ket{D_{5/2}}$.

\section*{Acknowledgments}
This research was supported by the National Research Foundation and the Ministry of Education of Singapore. A part of the calculations were carried out using 3TFLOP HPC Cluster at PRL, Ahmedabad.


\section*{Appendix A}

In this section we give details on the exact method followed to obtain the correction terms $\eta$ and $\zeta$. For our calculations, we follow the normalization convention for the reduced matrix elements $\langle I||T_{k}^{n}||I\rangle$ and $\langle\gamma^{\prime}J^{\prime}||T_{k}^{e}||\gamma J\rangle$ such that the Wigner-Eckert theorem takes the form
\begin{equation}
\braket{\gamma' j' m' | T_{k,q} | \gamma j m}=\frac{(-1)^{2k}}{\sqrt{2j + 1}}\braket{\gamma' j'|| T_{k} || \gamma j }\braket{j' m'|j m;k q}
\end{equation} where $T_q$ is the $q$ component of the spherical tensor of rank k, $T^{(k)}$, $j$,  $j'$ are angular momenta with $m$ and $m'$ their respective projections along the quantization axis. Note that from this convention it follows that
\begin{equation}
\braket{\gamma' j'||T_{k} || \gamma j } = (-1)^{(j-j')}\braket{\gamma j|| T_{k} || \gamma' j' }^*
\end{equation}which in our case, leads to a sign difference between the reduced matrix element and its conjugate. Applying the Wigner-Eckart theorem on the nuclear and electronic spaces we arrive at
\begin{align}
\label{HFI}
&\langle\gamma^{\prime}IJ^{\prime}F^{\prime}M_{F}^{\prime}|H_{HFI}|\gamma
IJFM_{F}\rangle = \delta_{F^{\prime}F}\delta_{M_{F}^{\prime}M_{F}}
(-1)^{I+J^\prime+F} \nonumber \\
&\times\sum_{k}\left\{
\begin{matrix}
I & J & F \\
J^\prime & I & k
\end{matrix}
\right\}  \,
\langle I||T_{k}^{n}||I\rangle\langle\gamma^{\prime}J^{\prime}||T_{k}^{e}||\gamma J%
\rangle,
\end{align} which is in agreement with the equation from \cite{Beloy2008} for the off diagonal matrix elements \footnote{The off diagonal matrix elements are defined in an unnumbered equation above Eq. (3) of \cite{Beloy2008}.}. Note that in an earlier work by the same authors there is a sign difference in  Eq. (3) of \cite{Beloy2008b} that differs by a factor of (-1)$^{(J-J^\prime)}$ from our result. The symbols represented by $I,J,F$ and $M_F$ are the nuclear spin, total angular momentum, hyperfine quantum number and its projection along the axis of quantization respectively. The correction terms are then derived for $\eta$ and $\zeta$ to be
\begin{eqnarray}
\eta  &=&\frac{(I+1)(2I+1)}{I}\mu^2\frac{|\langle D^0_{5/2}||T^e_1||D^0_{3/2}\rangle{|}^2}{E_{D^0_{5/2}} - E_{D^0_{3/2}}}, \\
\zeta  &=&\frac{(I+1)(2I+1)}{I}\sqrt{\frac{2I+3}{2I-1}} \nonumber \\
&&\times\frac{\mu{Q}\langle D^0_{5/2}||T^e_1||D^0_{3/2}\rangle\langle D^0_{5/2}||T^e_2||D^0_{3/2}\rangle}{E_{D^0_{5/2}} - E_{D^0_{3/2}}},
\end{eqnarray} where $\langle D^0_{5/2}||T^e_1||D^0_{3/2}\rangle$ and  $\langle D^0_{5/2}||T^e_2||D^0_{3/2}\rangle$ are given in Table \ref{error_matrix}. The values of the dipole moment $\mu$ \cite{Werth1995} are
\begin{equation}
\label{dipole}
\mu=0.937365(20)\,\mu_N
\end{equation} and the quadrupole moment $Q$
\begin{equation}
\label{quadrapole}
Q=0.235(3)\,\mathrm{b},
\end{equation} which comes from our D$_{3/2}$ hyperfine B constant given in Table \ref{hfc32} converted by the diagonal matrix element $\langle D^0_{3/2}|T^e_2|D^0_{3/2}\rangle$ from \cite{Sahoo2013}.  The coefficients in front of $\eta$ and $\zeta$ in Eqs.~(\ref{split1}), (\ref{split2}) and (\ref{split3}) can be found from
\begin{eqnarray}
c^{({W_F})}_\eta &=& \left(\left\lbrace\begin{matrix}
3/2 & 5/2 & F \\
3/2 & 3/2 & 1
\end{matrix}\right\rbrace\right)^2, \\
c^{({W_F})}_\zeta &=& \left\lbrace\begin{matrix}
3/2 & 5/2 & F \\
3/2 & 3/2 & 1
\end{matrix}\right\rbrace\left\lbrace\begin{matrix}
3/2 & 5/2 & F \\
3/2 & 3/2 & 2
\end{matrix}\right\rbrace.
\end{eqnarray}

\begin{table}[h]
\centering\caption{\label{error_matrix} Off diagonal elements of electronic spherical tensors of rank k (k$>$0) $T^e_k$.}
\begin{ruledtabular}
\begin{tabular}{p{4cm}|p{5cm}}
Elements & Value  \\
\hline
$\langle D^0_{5/2}||T^e_1||D^0_{3/2}\rangle$ & $995(10)\,\mathrm{MHz/\mu_N}$\\
$\langle D^0_{5/2}||T^e_2||D^0_{3/2}\rangle$ & $255(5)\,\mathrm{MHz/b}$ \\
\end{tabular}
\end{ruledtabular}
\end{table}

\section*{Appendix B}

In this appendix we discuss the model used for fitting the data.  This model incorporates the influence of hyperfine mixing between the $5D_{3/2}$ and $5D_{5/2}$ manifolds.  In $\mathrm{Ba}^{+}$ the fine structure splitting of the $5D$ level is very large ($24.0\,\mathrm{THz}$) and thus the influence of any mixing between the $5D_{3/2}$ and $5D_{5/2}$ manifolds can be determined by perturbation theory.

Since the Zeeman interaction only mixes states with the same $m_F$, we can restrict ourselves to a particular $m_F$.  Neglecting any mixing with the $D_{3/2}$ levels, the Hamiltonian in the presence of a magnetic field for the $m_F=0,\pm 1$ levels of $D_{5/2}$ is given by
\begin{align}
\label{hyperfine0}
H_0 = &\begin{pmatrix}
E_1 & 0 & 0 & 0 \\
0 & E_2 & 0 & 0 \\
0 & 0 & E_3 & 0 \\
0 & 0 & 0 & E_4
\end{pmatrix} \nonumber \\
+\,U_{m_F}^\dagger &\begin{pmatrix}
-\frac{3}{2} & 0 & 0 & 0 \\
0 & -\frac{1}{2} & 0 & 0 \\
0 & 0 & \frac{1}{2} & 0 \\
0 & 0 & 0 & \frac{3}{2}
\end{pmatrix}U_{m_F}(g_J - g_I)\mu_{B}B,
\end{align}
where $U_{m_F}$ is the unitary transformation between the $F$ and $IJ$ bases for the $m_F$ level of interest.  We note that we have omitted a term $g_I m_F \mu_B \mathbb{I}B$ which, being proportional to the identity matrix, simply adds to the energy eigenvalues and does not impact on any of the following discussion.

The Hamiltonian given in Eq.~(\ref{hyperfine0}) neglects mixing of the 5D$_{3/2}$ and 5D$_{5/2}$ levels. Strictly speaking, the hyperfine eigenstates are given by
\begin{eqnarray}
\ket{D_{5/2},F} &=& \alpha_{F}\ket{D^0_{5/2},F} + \beta_{F}\ket{D^0_{3/2},F},\\
\ket{D_{3/2},F} &=& \alpha_{F}\ket{D^0_{3/2},F} - \beta_{F}\ket{D^0_{5/2},F},
\end{eqnarray}
where the coefficients $\alpha_F, \beta_F$ are independent of $m_F$ and satisfy $\alpha_{F}^2 + \beta_{F}^2 = 1$. The full Zeeman Hamiltonian, $H_z=\mu_B(g_S S_z+g_L L_z+g_I I_z)B/\hbar$, then has the form
\begin{equation}
\begin{pmatrix}
H_{a} & H_{ab} \\ H_{ab}^{\dag} & H_{b}
\end{pmatrix}.
\end{equation}
where matrix elements of $H_a$, $H_b$, and $H_{ab}$ are given by
\begin{eqnarray}
\bra{D_{5/2},F}H_a\ket{D_{5/2},F'} &= \alpha_{F}\alpha_{F'}\bra{D^0_{5/2},F}H_z\ket{D^0_{5/2},F'} \nonumber\\
&  + \alpha_{F}\beta_{F'}\bra{D^0_{5/2},F}H_z\ket{D^0_{3/2},F'} \nonumber\\
&  + \beta_{F}\alpha_{F'}\bra{D^0_{3/2},F}H_z\ket{D^0_{5/2},F'} \nonumber\\
&  + \beta_{F}\beta_{F'}\bra{D^0_{3/2},F}H_z\ket{D^0_{3/2},F'}, \nonumber\\
\end{eqnarray}
\begin{eqnarray}
\bra{D_{5/2},F}H_{ab}\ket{D_{3/2},F'} &= \alpha_{F}\alpha_{F'}\bra{D^0_{5/2},F}H_z\ket{D^0_{3/2},F'} \nonumber\\
&  -\alpha_{F}\beta_{F'}\bra{D^0_{5/2},F}H_z\ket{D^0_{5/2},F'} \nonumber\\
& + \beta_{F}\alpha_{F'}\bra{D^0_{3/2},F}H_z\ket{D^0_{3/2},F'} \nonumber\\
& -\beta_{F}\beta_{F'}\bra{D^0_{3/2},F}H_z\ket{D^0_{5/2},F'}, \nonumber \\
\end{eqnarray}
\begin{eqnarray}
\bra{D_{3/2},F}H_b\ket{D_{3/2},F'} &= \alpha_{F}\alpha_{F'}\bra{D^0_{3/2},F}H_z\ket{D^0_{3/2},F'} \nonumber\\
& -\alpha_{F}\beta_{F'}\bra{D^0_{3/2},F}H_z\ket{D^0_{5/2},F'} \nonumber\\
& -\beta_{F}\alpha_{F'}\bra{D^0_{5/2},F}H_z\ket{D^0_{3/2},F'} \nonumber\\
& + \beta_{F}\beta_{F'}\bra{D^0_{5/2},F}H_z\ket{D^0_{5/2},F'}, \nonumber\\
\end{eqnarray}
The parameters $\beta_F$ can be determined from atomic structure calculations and to first order in the hyperfine interaction we have \cite{Beloy2008}
\begin{align}
\beta_F=&\frac{\langle D^0_{3/2},F|H_{HFI}|D^0_{5/2},F\rangle}{E_{D^0_{5/2}}-E_{D^0_{3/2}}}
= \frac{(-1)^{F+1}}{E_{D^0_{5/2}}-E_{D^0_{3/2}}} \nonumber \\
&\times\sum_{k}\left\{\begin{matrix}3/2 & 5/2 & F\\ 3/2 & 3/2 & k\end{matrix}\right\}\langle D^0_{3/2}||T^e_k||D^0_{5/2}\rangle \langle I||T^n_k||I\rangle.
\end{align}
Accuracy of a few percent can be obtained by including only the $k=1$ and $2$ terms in the summation. The reduced matrix elements $\langle I||T^n_k||I\rangle$ can be determined from the nuclear multipole moments and, from \cite{Beloy2008}, they are $2\sqrt{\frac{5}{3}} \mu$ and $\sqrt{5} Q$ for $k=1$ and $2$ respectively where $\mu$ and $Q$ are given in Eqs.~(\ref{dipole}) and (\ref{quadrapole}), respectively.  Using matrix elements $\langle D^0_{3/2}||T^e_1||D^0_{5/2}\rangle$ and $\langle D^0_{3/2}||T^e_2||D^0_{5/2}\rangle$ given in Table \ref{error_matrix} we obtain $\beta_1=0.915 \times 10^{-5}$, $\beta_2=1.478\times 10^{-5}$, and $\beta_3=1.698\times 10^{-5}$.

The terms in $H_{a}$ and $H_{b}$ proportional to $\alpha_{F}\alpha_{F}'\approx 1$ are simply the elements of the Zeeman Hamiltonian neglecting any mixing. All the other terms can be treated as a perturbation.  Elements of $H_{ab}$ only influence the energy levels at second order giving shifts $\sim(\mu_B B)^2/E_{FS}$ where $E_{FS}$ is the fine structure splitting.  For the $B$ fields considered in this work this amounts to shifts $\lesssim 0.25\,\mathrm{Hz}$ and thus we can neglect $H_{ab}$ altogether.  Moreover, the terms proportional to $\beta_F^2$ will contribute at most by $\sim \beta_F^2 \mu_B B$ which amounts to level shifts of only a few mHz.  Thus, concerning the measurements in the $D_{5/2}$ manifold, we need only to consider the terms in $H_a$ that are proportional to $\beta_F$ as a perturbation to the zero order Hamiltonian given in Eq.~(\ref{hyperfine0}).  The matrix elements $\bra{D^0_{5/2},F}H_z\ket{D^0_{3/2},F'}$ are all proportional to $(g_S-g_L)\mu_B B$ and the proportionality constants are given in Tables \ref{mf0} and \ref{mf1}.

\begin{table}[h!]
\centering\caption{\label{mf0} $\bra{D^0_{5/2},F}H_z\ket{D^0_{3/2},F'}$ elements scaled by $(g_s-g_L)\mu_B B$ for $m_F=0$.}
\begin{ruledtabular}
\begin{tabular}{c|cccc}
\diag{0em}{.8cm}{F}{F'} & 3 & 2 & 1 & 0 \\
\hline
1 & 0 & $\frac{1}{5\sqrt{5}}$ & 0 & $-\frac{1}{\sqrt{5}}$\\
2&$\frac{1}{5}\sqrt{\frac{3}{35}}$&0&$-\frac{1}{5}\sqrt{\frac{21}{5}}$&0\\
3&0&$-\frac{2}{5}\sqrt{\frac{6}{5}}$&0&0\\
4&$-2\sqrt{\frac{3}{35}}$&0&0&0\\
\end{tabular}\end{ruledtabular}\end{table}

\begin{table}[h!]
\centering\caption{\label{mf1} $\bra{D^0_{5/2},F}H_z\ket{D^0_{3/2},F'}$ elements scaled by $(g_s-g_L)\mu_B B$ for $m_F=\pm{1}$.}
\begin{ruledtabular}\begin{tabular}{c|cccc}
\diag{0em}{.8cm}{F}{F'} & 3 & 2 & 1 & 0 \\
\hline
1&0&$\frac{1}{10}\sqrt{\frac{3}{5}}$&$\mp\frac{3}{10}$&0\\
			2&$\frac{2}{5}{\sqrt{\frac{2}{105}}}$&$\mp\frac{1}{10}\sqrt{\frac{7}{3}}$&$-\frac{3}{10}\sqrt{\frac{7}{5}}$&0\\
			3&$\mp\frac{1}{5\sqrt{6}}$&$-\frac{8}{5\sqrt{15}}$&0&0\\
			4&$-\sqrt{\frac{3}{14}}$&0&0&0\\
\end{tabular}
\end{ruledtabular}
\end{table}

For the $m_F=0$ case, the perturbation has the form
\begin{widetext}
\begin{equation}
\begin{pmatrix}
0 & \frac{1}{5\sqrt{5}}\beta_2 - \frac{1}{5}\sqrt{\frac{21}{5}}\beta_1 & 0 & 0 \\
\frac{1}{5\sqrt{5}}\beta_2 - \frac{1}{5}\sqrt{\frac{21}{5}}\beta_1 & 0 & -\frac{2}{5}\sqrt{\frac{6}{5}}\beta_2 +\frac{1}{5}\sqrt{\frac{3}{35}}\beta_3 & 0 \\
0 & -\frac{2}{5}\sqrt{\frac{6}{5}}\beta_2 +\frac{1}{5}\sqrt{\frac{3}{35}}\beta_3 & 0 & -2\sqrt{\frac{3}{35}}\beta_3 \\
0 & 0 & -2\sqrt{\frac{3}{35}}\beta_3 & 0 \\
\end{pmatrix}.
\end{equation}
\end{widetext}
Due to the fact that only $F=3$ and $F=4$ levels are mixed significantly by the Zeeman interaction, the unitary transformation that diagonalizes Eq.~(\ref{hyperfine0}) has the approximate form
\begin{equation}
\begin{pmatrix}
\mathbb{I} & 0 \\ 0 & $R$
\end{pmatrix}
\end{equation}
where $R$ is a rotation matrix that depends on the strength of the magnetic field. Consequently, in the basis of states that diagonalizes Eq.~(\ref{hyperfine0}), there are no significant diagonal elements of the perturbation associated with the states $|1,0\rangle$ and $|2,0\rangle$. Calculations confirm shifts of $<1\,\mathrm{Hz}$ for magnetic fields of $<2\,\mathrm{G}$.  Thus, for the energies of these two states we can neglect mixing with the $D_{3/2}$ manifold altogether.  We note that the lack of diagonal elements in the perturbation for the $m_F=0$ case is a consequence of the magnetic dipole selection rule that $m_F'=0\leftrightarrow m_F=0$ is forbidden when $\Delta F=0$.  Consequently, our previous measurements for the $D_{3/2}$ level were not affected by hyperfine mixing of the fine structure levels.

For the  $m_F = \pm 1$ case, the perturbation has the form
\begin{widetext}
\begin{equation}
\begin{pmatrix}
\mp\frac{3}{5}\beta_1 & -\frac{3}{10}\sqrt{\frac{7}{5}}\beta_1 + \frac{1}{10}\sqrt{\frac{3}{5}}\beta_2 & 0 & 0 \\
-\frac{3}{10}\sqrt{\frac{7}{5}}\beta_1 + \frac{1}{10}\sqrt{\frac{3}{5}}\beta_2 & \mp \frac{1}{5}\sqrt{\frac{7}{3}}\beta_2 & -\frac{8}{5}\frac{1}{\sqrt{15}} \beta_2 + \frac{2}{5}\sqrt{\frac{2}{105}}\beta_3 & 0 \\
0 & -\frac{8}{5}\frac{1}{\sqrt{15}}\beta_2 + \frac{2}{5}\sqrt{\frac{2}{105}}\beta_3 & \mp \frac{2}{5}\frac{1}{\sqrt{6}}\beta_3 & -\sqrt{\frac{3}{14}}\beta_3 \\
0 & 0 & -\sqrt{\frac{3}{14}}\beta_3 & 0
\end{pmatrix}.
\end{equation}
\end{widetext}
The states of interest here are those associated with the $F=3$ and $F=4$ levels. From the approximate form of the unitary transformation that diagonalizes Eq.~(\ref{hyperfine0}), these levels are only influenced by the terms proportional to $\beta_3$, and we note that these elements provide level shifts on the order of $\beta_3 \mu_B B/\hbar \sim 2\pi\times 25\,\mathrm{Hz}$.  For the purposes of modeling the energy levels of interest we therefore use Eq.~(\ref{hyperfine0}) for the $m_F=0$ states measured, while for the $F=3$ and $F=4$ levels we include the perturbation
\begin{equation}
\begin{pmatrix}
0 & 0 & 0 & 0 \\
0 & 0 & \frac{2}{5}{\sqrt{\frac{2}{105}}} & 0 \\
0 & \frac{2}{5}{\sqrt{\frac{2}{105}}} & \mp\frac{2}{5\sqrt{6}} & - \sqrt{\frac{3}{14}} \\
0 & 0 &  -\sqrt{\frac{3}{14}} & 0
\end{pmatrix}\beta_3(g_s-g_L)\mu_B B,
\end{equation} in Eq.~(\ref{hyperfine0}).

\section*{Appendix C}

\begin{table}[h]
\centering\caption{\label{hfc32}D$_{3/2}$  Hyperfine coupling constants.}
\begin{ruledtabular}
\begin{tabular}{lrrr}
            & A\,(Hz)           & B\,(Hz)          & C\,(Hz) \\
\hline Uncorr.       & 189730524.90(32) & 44538793.7(10) & 32.465(44) \\
$\eta$ corr.  &          805(16) &      $-$1610(32) & $-$ \\
$\zeta$ corr. &      164.2(42) &        411(10) & $-$2.933(75) \\
Corr.         &    189731494(17) &    44537594(34) & 29.533(86)\\
\end{tabular}
\end{ruledtabular}
\end{table}

In this section we review the hyperfine constants for the D$_{3/2}$, which are presented in Table \ref{hfc32}. These hyperfine constants differ with respect to the most recently published values in \cite{Lewty2013}. This is due to previously having an error in the code used in the calculations of the off diagonal elements given in Table \ref{error_matrix} compared with \cite{Lewty2012} and they also differ due to a factor of $(-1)^{J-J'}$ error we made with correction factor $\eta$ for which we apologize.

\bibliography{octupole.bib}

\end{document}